\documentclass[11pt]{article}
\usepackage{jcappub}
\usepackage{epsfig}
\usepackage{amssymb,amsmath,bbm}
\usepackage{graphicx}
\usepackage{verbatim}
\usepackage{appendix}
\usepackage{color}

\newcommand{\X}{\vec{x}}

\newcommand{\Pl}{M_\mathrm{Pl}}
\newcommand{\e}{\mathrm{e}}

\title{A general method for the construction of stable Galileon models consistent with the Planck data results}
\author[a]{Rio Saitou,}
\author[b]{Emilio Elizalde}
\affiliation[a]{Institute for Cosmic Ray Research, The University of Tokyo, 5-1-5 Kashiwa-no-ha, Kashiwa City, Chiba 277-8582 (Japan)}
\affiliation[b]{Consejo Superior de Investigaciones Cientificas, ICE-CSIC and IEEC, UAB Campus, 08193 Bellaterra, Barcelona (Spain)}
\emailAdd{rio.saitou@ipmu.jp}
\emailAdd{elizalde@ieec.uab.es}
\date{\today}

\abstract{The reconstruction procedure, which has proven quite useful to obtain viable models of the universe evolution, is here employed in order to construct inflation models. It has the advantages that it ensures full consistency with astronomical observations and that it allows to evaluate the stability of the resulting cosmological model quite easily. The reconstruction for two different types of Lagrangian, included in the frame of G-inflation, is carried out in detail and explicit models for each Lagrangian are constructed. As a bonus for having used this reconstruction formalism, the final models are easily adjusted to satisfy the observational constraints---imposed by the most recent data releases of the Planck mission---on the spectral index, the tensor to scalar ratio, and the running of the spectral index. Further, it turns also to be not difficult to make the models stable. Thus, the method here developed provides a general and very efficient tool, a quite natural procedure to construct models consistent with very precise observations. It can also be applied to other  models, besides the ones here considered.
}


\begin{document}
	

\maketitle

\section{Introduction}

Inflation in the early universe \cite{Starobinsky:1979ty, Linde:1981mu, as1} has become a part of the standard cosmological model. As is known, its role is fundamental for several important reasons, starting with its role in the amplification of the incipient quantum density or curvature fluctuations. One problem, as of now, is that there are too many models of inflation and one should try to establish which is the one that better describes the very first stages of the evolution of our universe. Actually, the true nature of inflation is becoming more and more clear thanks to the latest data releases corresponding to the very precise observations of the cosmic microwave background (CMB) carried out recently by the Planck satellite \cite{Ade:2015lrj, Ade:2015xua}, the BICEP2 experiment \cite{Ade:2014xna} and the Wilkinson Microwave anisotropy probe (WMAP) \cite{Spergel:2003cb}. The most common mechanism driving inflation involves a scalar field and relies on the specific form of the potential of this scalar field, called the inflaton \cite{Linde:1981mu, as1, Linde:1983gd}.

Since the appearance of this potential-driven inflation, a good number of novel models have been devised by modifying the kinetic term of the theory. These include k-inflation \cite{ArmendarizPicon:1999rj}, ghost inflation \cite{ArkaniHamed:2003uz}, Dirac-Born-Infeld inflation \cite{Alishahiha:2004eh}, the simplest higher-derivative theory without ghosts, which is modified Gauss-Bonnet gravity \cite{nos05}, G-inflation \cite{Kobayashi:2010cm}, and generalized G-inflation \cite{Kobayashi:2011nu}. Some of these models are already on the verge of being excluded by the observational constraints obtained from the Planck and WMAP data releases, but many of them are still compatible with these astronomical observations and thus one cannot yet decide which is the preferred inflationary model.

As an alternative approach in order to build inflation models, the Hamilton-Jacobi formalism of inflation has been put forward. It also deals with a canonical scalar field model with a potential \cite{Salopek:1990jq} but in this formalism one provides the Hubble parameter as a function of the scalar field by hand, and one can then determine the potential term from it. This formalism has been extended to models with a non-canonical kinetic term \cite{Gauthier:2008mq, Bean:2008ga}.
In parallel with this development, but now mainly in the context of the present acceleration of the universe, a formalism which is quite similar to this one, termed the reconstruction method, was pioneered in \cite{Nojiri:2006gh} and subsequently developed with great success (for a review of cosmological reconstruction, see \cite{no11r, no06r}). In the reconstruction formalism, using re-scaling of the scalar field, one can eventually provide the Hubble parameter as a function of time instead of the scalar field, so that it becomes easier to compare the model with the observations than in the Hamilton-Jacobi formalism. So far, however, the reconstruction formalism has been used in this context in a limited number of models only \cite{Matsumoto:2010uv, Shirai:2012iw, Saitou:2011hv}.

The purpose of the present work is to explicitly develop and carry to the end the reconstruction formalism for two different types of Lagrangian, included in the frame of G-inflation, for which the reconstruction formalism has not been thoroughly devised. The specific form in one of them consists of a generic function of the kinetic term and a potential term of the scalar field. The other one exhibits the same form as G-inflation \cite{Kobayashi:2010cm}. Once we finish with the construction of concrete models for each Lagrangian, we will provide a proper Hubble parameter together with a quantity related to the sound speed of the scalar field and demonstrate that they satisfy the stringent observational constraints imposed on the spectral index, on the tensor to scalar ratio, and on the running of the spectral index, by the most recent data releases of the Planck mission \cite{Ade:2015lrj}. Further, as we will see in what follows, it will also be not difficult to make the models stable. We will see how the method here developed provides a general, efficient and quite natural procedure to construct models which are consistent with the most precise observations. We will do that by working explicitly with some examples but the method can also be applied to other models besides the ones here considered.

The  paper is organized as follows. In Sec.~\ref{Review}, we give a brief review of
Galileon cosmology. In Sec.~\ref{Reconstruction}, we describe the reconstruction formalism for the two types of Lagrangian considered and construct specific models for each of the Lagrangians. Sec.~\ref{Conclusions} is devoted to conclusions. Finally, in an Appendix, we describe the reconstruction formalism for the case of exact de Sitter background solutions in G-inflation.

\section{Short review of Galileon cosmology}
\label{Review}

In this section we provied a summary of the Galileon theory \cite{Deffayet:2009mn, Deffayet:2010qz} and its cosmological consequences following the notations in \cite{Kobayashi:2011nu}.
We actually consider only a part of the Galileon action, namely
\begin{equation}
\label{S}
S = \int d^4x \sqrt{-g}\left( \frac{\Pl^2}{2}R + K(\phi, X) - G(X)g^{\mu\nu}\nabla_\mu\nabla_\nu\phi \right) \ ,
\end{equation}
where $\Pl=1.2\times10^{19}\text{GeV}$ is the Planck mass, $R$ the scalar curvature,
$K$ and $G$ are generic functions of a single scalar field, $\phi$, and $X\equiv -\frac{1}{2}g^{\mu\nu}\nabla_\mu\phi\nabla_\nu\phi$.
We restrict the function $G$  to depend on $X$, only, and fix other Galileon terms, as
$G_4=\Pl^2/2$ and $G_5=0$, just for simplicity.

For describing the background evolution of the isotropic homogeneous
universe, we assume that $\phi =\phi(t)$ and work with the flat FLRW metric $ds^2 = -dt^2 + a(t)^2d\X^2$, where $a(t)$ is the scale factor.
The Einstein equations and the field equation derived from the action (\ref{S}), in this background, are
\begin{align}
\label{bg}
 &3\Pl^2H^2 = 2XK_X-K +6X\dot{\phi} HG_X \ ,\\
 &\Pl^2(3H^2+ 2\dot{H}) =  -K + 2X\ddot{\phi}G_X \ , \\
 &(\dot{\phi}K_X +6HXG_X)^{\cdot} + 3H(\dot{\phi}K_X +6HXG_X) = K_\phi \ ,
\end{align}
where $H \equiv a/\dot{a}$ is the Hubble parameter, the dot meaning time derivative. The indexes on the functions $K$ and $G$ denote corresponding partial derivatives. These equations determine the background evolution of the universe.

Next, we will obtain the quadratic action for the scalar and tensor modes of the cosmological perturbations of the full action (\ref{S}). We fix the unitary gauge $\phi =\phi(t)$ and perturb the metric using the ADM formalism, in the form
\begin{equation}
\label{ }
ds^2 = -{\cal N}^2dt^2 + \gamma_{ij}(dx^i +N^idt)(dx^j +N^jdt) \ ,
\end{equation}
where
\begin{equation}
\label{ }
{\cal N} = 1+\alpha, \ N_i = \partial_i \beta,\ \gamma_{ij} = a^2(t)\mathrm{e}^{2\zeta}\left( \delta_{ij}
+ h_{ij} + \frac{1}{2}h_{ik}h_{kj}\right) \ .
\end{equation}
In the above, $\alpha$ and $\beta$ are the auxiliary fields for the scalar perturbations, $\zeta$ for the curvature perturbations and $h_{ij}$ for the tensor perturbations, which satisfy $\partial_ih_{ij} = 0$ and $h_{ii}=0$. After removing the auxiliary fields by solving the constraint equations, we obtain the following quadratic actions for the tensor and for the curvature perturbations, respectively,
\begin{align}
\label{S2}
    S_T^{(2)}&= \frac{\Pl^2}{8}\int dtdx^3 a^3\left[ \dot{h}_{ij}^2 - \frac{(\vec{\nabla}h_{ij})^2}{a^2} \right] \ ,  \nonumber \\
    S_{\zeta}^{(2)}&= \int dtdx^3 a^3\left[ {\cal G}_S\dot{\zeta}^2 - {\cal F}_S\frac{(\vec{\nabla}\zeta)^2}
    {a^2} \right] \ ,
\end{align}
where
\begin{align}
\label{F}
    &{\cal F}_S= \Pl^4\left(\frac{H}{\Theta} - \frac{\dot \Theta}{\Theta^2}\right) - \Pl^2 \ ,  \\
    \label{G}
    &{\cal G}_S= \Pl^4\frac{\Sigma}{\Theta^2} +3\Pl^2 \ , \\
    &\Theta \equiv -\dot{\phi}XG_X + \Pl^2H \ , \nonumber \\
    &\Sigma \equiv XK_X +2X^2K_{XX} +12H\dot \phi XG_X+6H\dot\phi X^2G_{XX} - 3\Pl^2H^2\ . \nonumber
\end{align}

The square of the sound speed of the tensor perturbations is the same as the light speed,
and for the case of the curvature perturbations, we have
\begin{equation}
\label{cs}
c_s^2= \frac{{\cal F}_S}{{\cal G}_S}\ .
\end{equation}
We need ${\cal F}_S>0$ and ${\cal G}_S>0$ in order to stabilize the system against the ghost and the gradient instabilities. Furthermore, to prohibit superluminal propagation,
$c_s^2>1$, what is compulsory for avoiding the closed time-path problem, we require ${\cal G}_S-{\cal F}_S\geq0$.
To summarize, for both stability and non-superluminal propagation, we need to impose
\begin{equation}
\label{SC}
{\cal F}_S > 0 \quad\text{and}\quad {\cal G}_S-{\cal F}_S\geq0\ .
\end{equation}

To obtain the tensor and the scalar power spectrums, we assume that the following parameters are nearly constant:
\begin{equation}
\label{SV}
\epsilon \equiv -\frac{\dot H}{H^2} \simeq \text{const},\quad f_s \equiv \frac{\dot{{\cal F}_S}}{H{\cal F}_S}
\simeq \text{const},\quad g_s \equiv \frac{\dot{{\cal G}_S}}{H{\cal G}_S} \simeq \text{const}\ .
\end{equation}
Here, we further assume that all parameters (\ref{SV}) are much smaller than 1, namely $\epsilon, f_s, g_s \ll 1$
\footnote{In \cite{Kobayashi:2011nu},  the power spectra of the perturbations and the spectral index were derived {\it without} the assumption that $\epsilon, f_s, g_s \ll 1$. Our expressions for the power spectra and the spectral index do match those obtained in \cite{Kobayashi:2011nu}, after imposing these conditions to the corresponding expressions.}.
After deriving the mode functions from the actions (\ref{S2}),
the tensor and scalar spectra are obtained, up to first order in the parameters (\ref{SV}), as
\begin{align}
\label{PS}
&{\cal P}_h = \left. \frac{2}{\Pl^2}\frac{H^2}{\pi^2}\right|_{k = aH} \ , \\
\label{SS}
&{\cal P}_\zeta = \left. \frac{H^2}{8\pi^2c_s{\cal F}_S}\right|_{c_sk = aH} \ ,
\end{align}
where $k$ is a co-moving wave number of the perturbations.
It is apparent that the tensor spectrum depends on time only through the Hubble parameter.
On the other hand, the scalar spectrum depends on time not only through the Hubble parameter
but also through the sound speed of the scalar perturbations and the quantity ${\cal F}_S$.
The spectral index, up to first order in the parameters (\ref{SV}), is
\begin{equation}
\label{obs1}
n_s -1 = -2\epsilon - \frac{3}{2}f_s + g_s \ ,
\end{equation}
and the tensor to scalar ratio reads
\begin{equation}
\label{obs2}
r = 16\frac{{\cal F}_S}{\Pl^2}c_s \ .
\end{equation}
The running of the spectral index is provided by
\begin{equation}
\label{obs3}
\alpha_s = -\frac{1}{H}\left( 2\dot \epsilon +\frac{3}{2}\dot f_s - \dot g_s \right) \ .
\end{equation}
%
%
The observational values for these quantities, at the pivot scale $k_*= 0.05\mathrm{Mpc}^{-1}$, reported by the Planck mission \cite{Ade:2015lrj} are
\begin{align}
\label{obs}
   \mathrm{ln}(10^{10}{\cal P}_\zeta) &= 3.062\pm0.029 \ , \nonumber \\
    ns &= 0.9690\pm 0.0063 \ , \nonumber \\
    r&< 0.176 \ , \nonumber \\
    \alpha_s&= -0.0076^{+0.0092}_{-0.0080},\  \text{(Planck TT+lowP+lensing)}\ .
\end{align}
Let us now imagine that we have obtained an explicit function of time for $H$ and a quantity related to $c_s$ from the astronomical observations data, for example, from (\ref{PS}) and (\ref{SS}). An action which will produce both this function $H$ and the quantity related to $c_s$ will be obviously consistent with the observation. The reconstruction method enables us to construct such an action. Using this method one is able in fact to provide an explicit function of time for $H$ and the quantity related to $c_s$, which are consistent with observations, and the stability condition (\ref{SC})
{\textit{by hand}}, as inputs of the model we consider, and by using them one can derive an action which will produce those inputs as the result of the evolution of the system. In the next section, we will formulate the reconstruction procedure for two different types of Lagrangian.

\section{Reconstruction procedure}
\label{Reconstruction}
In this section, we formulate the reconstruction of the action for two different types of Lagrangian
with an explicit function of $H$ and a quantity related to $c_s$.

To start, we recall the e-folding number definition,
\begin{align}
\label{ }
&N = \int^{t_e}_t Hdt = \mathrm{ln}\frac{a_e}{a}\ , \\
&\frac{d}{dt} = -H\frac{d}{dN} \ , \nonumber
\end{align}
where $t_e$ and $a_e$ are the cosmic time and the scale factor when $N=0$, respectively.
We use $N$ as affine parameter, instead of $t$, because we will construct
models of inflation, and in such models $N$ is a more useful parameter to directly constrain model parameters by observations, than is $t$. We assume that inflation ended at $N=0$.
At the same time, we redefine the scalar field as
\begin{align}
\label{fr}
&\phi = -\int \frac{\sqrt{2q(\varphi)}}{H(\varphi)}\partial \varphi \ , \nonumber \\
&X = -\frac{1}{2}g^{\mu\nu}\partial_\mu \phi \partial_\nu \phi
   = -\frac{q(\varphi)}{H^2(\varphi)}g^{\mu\nu}\partial_\mu \varphi \partial_\nu \varphi \ ,
\end{align}
and we assume that we can take $\varphi = N$ as one of the background solutions
of the system just by choosing a field redefinition (\ref{fr}) properly. We note that the function
$q(\varphi=N)$ must be positive, since we consider a real scalar field theory. In this case,
at the background level, we get following relations
\begin{align}
\label{re}
&\phi = -\int \frac{\sqrt{2q(N)}}{H(N)}\partial N \ , \nonumber \\
&X = \frac{1}{2}\dot \phi^2  = q(N) \ , \
\dot \phi = \sqrt{2q(N)} \ .
\end{align}
The background equations (\ref{bg}) using the above relations (\ref{re}) become
\begin{align}
\label{bg1}
   &3\Pl^2H(N)^2 = 2qK_{q}(N, q) - K(N, q) +6H\sqrt{2q^3}G_{q}(q)  \ , \nonumber \\
   &\Pl^2(3H(N)^2-2HH'(N)) = -K(N,q)  -Hq'\sqrt{2q}G_{3q}(q) \ , \nonumber \\
    &(\sqrt{2q}K_{q}(N,q) + 6HqG_{q}(q))' -3(\sqrt{2q}K_{q}(N,q)+6HqG_{q}(q)) =\frac{K_{N}(N,q)}{\sqrt{2q}} \ ,
\end{align}
where the prime denotes a total derivative with respect to $N$, and the functions with
the indices $N$ or $q$ mean partial derivatives with respect to those indices.
Next, we will formulate the reconstruction explicitly for two different types of Lagrangian.
\subsection{$P(X)-V(\phi)$}

Consider the following Lagrangian
\begin{equation}
\label{L1}
K(\phi, X) = P(X) - V(\varphi(\phi)) \ , \ G(X) = 0 \ .
\end{equation}
The background equations (\ref{bg1}) for this Lagrangian are
\begin{align}
\label{bg2}
    &3\Pl H^2 = 2qP_q-P+V  \ , \\
    \label{2bg2}
    &\Pl(3H^2-2HH') = -P+V  \ , \\
    \label{f2}
    &(\sqrt{2q}P_q)' -3\sqrt{2q}P_q = -\frac{V_N}{\sqrt{2q}}= -\frac{V'}{\sqrt{2q}}\ .
\end{align}
The total derivative of the equation (\ref{2bg2}) with respect to $N$ becomes
\begin{equation}
\label{ }
V' = q'P_q + 6qP_q -2\Pl(H'^2+HH'') \ .
\end{equation}
The quantities (\ref{F}), (\ref{G}) and the square of sound speed (\ref{cs}) read
\begin{align}
\label{ }
&{\cal F}_S = \Pl^2 \epsilon = \frac{qP_q}{H^2}, \quad
{\cal G}_S = \frac{q(P_q+2qP_{qq})}{H^2} \ , \nonumber \\
&c_s^2 = \frac{P_q}{P_q + 2qP_{qq}} \ ,
\end{align}
where we have used the equations (\ref{bg2}) and (\ref{2bg2}) for ${\cal F}_S$.
Substituting them into the field equation (\ref{f2}), we obtain
\begin{align}
\label{I}
&q'P_q = P' = \Pl^2\cdot \frac{2c_s^2}{c_s^2+1}(H'^2+HH'') \equiv I'(N)\ , \nonumber \\
&P(N) = I(N) + C \ ,
\end{align}
where $C$ is an integration constant.
If we provide explicit functions for $H(N)$ and $c_s(N)$,
we can obtain the function $I(N)$ as an explicit function of $N$ after integrating (\ref{I}).
Substituting (\ref{I}) into the background equations (\ref{bg2}) and (\ref{2bg2}), we get
\begin{align}
\label{J}
    &\frac{q'}{q} = \frac{I'}{\Pl^2HH'} \equiv J'(N) \ , \nonumber  \\
    &\mathrm{ln}\frac{q}{q_0} = J(N) \ ,
\end{align}
where $q_0$ is a constant which has the same mass scale as $q$.
Then, if we can invert the function $J(N)$ as $N(q) = J^{-1}[\mathrm{ln}(q/q_0)]$,
we arrive to
\begin{align}
\label{P}
    P(q) &= I\circ J^{-1}[\mathrm{ln}(q/q_0)] + C \ , \\
    \label{V}
    V(N) &= I(N) + C + \Pl^2 (3H^2(N) -2HH'(N)) \ .
\end{align}
Here, we will set $C=0$, since $C$ always vanishes from the Lagrangian $P-V$.
By plugging now $q$ and $N$ into $X$ and $\varphi$ and using the field redefinition (\ref{re}), we finally obtain the form of the original Lagrangian, as
\begin{align}
\label{L}
P(X)-V(\varphi(\phi)) &= I\circ J^{-1}[\mathrm{ln}(X/q_0)] \nonumber \\
                               &- I(\varphi(\phi)) -\Pl^2 \left(3H^2(\varphi(\phi)) - 2HH_\varphi(\varphi(\phi))\right)  \ .
\end{align}
The Lagrangian (\ref{L}) created by using this formalism\footnote{A different formalism for the reconstruction of this Lagrangian can be found in \cite{Gauthier:2008mq, Bean:2008ga}. These authors use another affine parameter, instead of the e-folding number $N$.} will produce $H(N)$ and $c_s(N)$ which we have provided at (\ref{I}). If we tune $H(N)$ and $c_s(N)$ so as to satisfy the observational constraints (\ref{obs}) and the stability condition (\ref{SC}), the reconstructed Lagrangian fulfills these constraints automatically. And if we use this formalism, we can ensure the stability
of the system from the start, a very important property which cannot be generically ensured by using other methods. Hence, with our procedure it becomes relatively much easier to construct stable models than with other techniques.

We shall now make a comment on the specific case $V=0$. In this case, we obtain
\begin{equation}
\label{ }
I' = -\Pl^2 (6HH' -2H'^2- 2HH'')
\end{equation}
from (\ref{V}), and substituting this into (\ref{I}), we get
\begin{equation}
\label{ck}
c_s^2 = -1 + \frac{H'}{3H} + \frac{H''}{3H}\ .
\end{equation}
If $V=0$, $c_s$ and $H$ are not independent from each other and, therefore, we need just to provide either $c_s(N)$ or $H(N)$ in order to determine the unknown function $P(X)$, using the same procedure as (\ref{J})-(\ref{P}). This exactly corresponds to a previous result by \cite{Matsumoto:2010uv}.

\subsubsection{Example 1}

We will now consider a specific model for the Lagrangian $P(X)-V(\phi)$.
Let us consider two explicit functions for $H(N)$ and $c_s(N)$, namely
\begin{align}
\label{H1}
    H(N)&= H_0(4N+p)^{\frac{p}{4}}  \ ,  \\
    \label{cs1}
    c_s(N)&= \sqrt{\frac{1}{3}} = \text{const.}\ ,
\end{align}
where $p$ and $H_0$ are constants, to be determined by observations.
The values of ($\epsilon,\ f_s,\ g_s$) for this model are
\begin{equation}
\label{ }
\epsilon = \frac{p}{4N+p},\quad f_s = g_s = \frac{4}{4N+p}\ ,
\end{equation}
and the values of the observables (\ref{obs1})-(\ref{obs3}),
\begin{equation}
\label{ }
n_s-1 = -\frac{2p+4}{4N+p},\quad r = \frac{16p}{\sqrt{3}(4N+p)}, \quad \alpha_s = -\frac{8p+16}{(4N+p)^2} \ .
\end{equation}
The tensor to scalar ratio $r$ is smaller by a factor $\sqrt{3}$ than the value when $c_s^2=1$.
If we take $N=60$ as the time when the curvature perturbation whose co-moving wave number is 
the pivot scale $k_* = 0.05\mathrm{Mpc}^{-1}$ grew up to the scale $c_s/H|_{N=60}$, 
the observable (\ref{obs}) provides the following constraint on $p$: $1< p < 2.5$.
Hereafter, we will always take $N=60$ as the time when the curvature perturbation whose co-moving wave number is the pivot scale $k_* = 0.05\mathrm{Mpc}^{-1}$ grew up to the scale $c_s/H|_{N=60}$.
The stability condition (\ref{SC}) becomes
\begin{align}
\label{SC1}
&{\cal F}_S = \Pl^2\frac{p}{4N+p}>0\ , \nonumber \\
&{\cal G}_S-{\cal F}_S = \left( \frac{1}{c_s^2}-1\right){\cal F}_S= 2{\cal F}_S\geq0 \ .
\end{align}
For $0\leq N\leq60$, $p>0$ or $p<-240$ is required from (\ref{SC1}).
Combining these constraints, the parameter $p$ must satisfy
\begin{equation}
\label{ }
1<p<2.5,
\end{equation}
during inflation.
On the other hand, $H_0$ is also determined from the observational value of (\ref{SS}) coming from the Planck data, as
\begin{equation}
\label{ }
H_0 = \Pl \frac{2\pi\sqrt{2p}}{3^{1/4}(4\cdot 60+p)^{\frac{p+2}{4}}}{\cal P}_\zeta^{1/2}  \ .
\end{equation}

Now, we reconstruct the Lagrangian using the two explicit functions (\ref{H1}) and (\ref{cs1}).
Using (\ref{I}) and (\ref{J}), we find
\begin{align}
\label{P1}
&P = \frac{\Pl^2}{2}HH' \ ,
\end{align}
and
\begin{align}
\label{q1}
    \mathrm{ln}\frac{q}{q_0} &= \frac{1}{2}\mathrm{ln}\left( \frac{HH'}{H_0^2}\right) \nonumber \\
                                   &= \frac{1}{2}\mathrm{ln}\left( p(4N+p)^{\frac{p-2}{2}}\right) \ .
\end{align}
We can fix the integration constant in $J(N)$ to $H_0^2$ without loss of generality.
If we take $p=2$, $q$ becomes a constant and we cannot invert about $N$.
For this model, nevertheless, we can reconstruct the kinetic function $P(q)$ without inverting $q$ in $N(q)$
for all values of $p$, since both $q$ and $P$ are functions of $HH'$. Solving $HH'$ about $q$ and substituting
it into (\ref{P1}), we obtain
\begin{equation}
\label{ }
P(q) = \frac{\Pl^2H_0^2}{2}\left(\frac{q}{q_0}\right)^2 \ .
\end{equation}
On the other hand, the potential term becomes
\begin{align}
\label{V1}
    V
      &= 3\Pl^2H_0^2(4N+p)^{\frac{p}{2}}\left( 1 -\frac{\epsilon}{2}\right)  \ .
\end{align}
The field redefinition (\ref{re}) for this model yields
\begin{align}
\label{ }
\phi 
       &= \frac{2p^{1/4}\sqrt{q_0}}{H_0(p-6)}(4N+p)^{-\frac{p-6}{8}} \ ,
\end{align}
and substituting this into (\ref{V1}), we obtain
\begin{align}
\label{ }
&V(\phi) = 3\Pl^2H_0^2\left( \frac{H_0(p-6)}{2p^{1/4}\sqrt{q_0}}\phi\right)^{-\frac{4p}{p-6}}
              \left( 1 -\frac{\epsilon(\phi)}{2}\right) \ , \nonumber \\
&\epsilon(\phi) \equiv p\left( \frac{H_0(p-6)}{2p^{1/4}\sqrt{q_0}}\phi\right)^{\frac{8}{p-6}} \ .
\end{align}
Inserting $q$ into $X$ and fixing the integration constant as $q_0 = \Pl H_0 M^2/\sqrt{2}$, where $M$ is a constant, we finally obtain the reconstructed Lagrangian
\begin{equation}
\label{ }
P(X)-V(\phi) = \frac{X^2}{M^4} - 3\Pl^2 H_0^2\left( \frac{p-6}{\sqrt{2}p^{1/4}\Pl M}\phi\right)^{-\frac{4p}{p-6}}\left( 1 -\frac{\epsilon(\phi/M)}{2}\right) \ .
\end{equation}
This model does not have a canonical kinetic term since we have provided the constant sound speed
$c_s\neq 1$, and the potential term is a power of the scalar field and a monomial, if we ignore the slow-roll parameter in the potential. Moreover, from (\ref{q1}), the kinetic term is proportional to $H^2\epsilon$. As in (\ref{V1}), the potential term is roughly proportional to the Hubble parameter; $V\sim 3\Pl^2H^2$ and, therefore, this model inflates the universe through slow-rolling of the scalar field, as expected from the form of the Lagrangian (\ref{L1}).
We note that the constant $M$ does not affect the evolution of the system since we can remove it from the Lagrangian by defining a dimensionless scalar field as $\hat{\phi}\equiv \phi/M$. We need some mechanism for the reheating process after inflation ends. There are some works about reheating with  non-canonical kinetic terms by gravitational particle production (see, e.g., \cite{Ford:1986sy}), but a detailed analysis of the reheating stage is beyond the scope of the present paper.

If, in particular, we take $p=2$  and ignore the slow roll parameter $\epsilon$ in the potential,
we obtain a quadratic potential
\begin{equation}
\label{ }
V^{quadra}(\phi) \simeq \frac{12\Pl H_0^3}{M^2}\phi^2 \ .
\end{equation}
The value $p=2$ is consistent with the astronomical observations and with the stability condition, and hence this inflation model with quadratic potential can {\textit{survive}} if we modify the kinetic term of the model properly, such as we have done. Of course, the model we have considered here is just an example and we can easily construct other models in the same fashion which are stable and consistent with observations by using this powerful reconstruction formalism.

\subsection{G-inflation}

In this subsection, we consider a different Lagrangian, namely
\begin{equation}
\label{L2}
K(\phi, X) = P(X)  \ , \ G(X) = G(X) \ .
\end{equation}
The background equations (\ref{bg1}) for this Lagrangian become
\begin{align}
\label{}
    &3\Pl H^2 = 2qP_q-P + 6H\sqrt{2q^3}G_q  \ , \\
    &\Pl(3H^2-2HH') = -P -H\sqrt{2q}q'G_q  \ , \\
    \label{f3}
    &(\sqrt{2q}P_q+6HqG_q)' -3(\sqrt{2q}P_q+6HqG_q) = 0
\end{align}
and this field equation tells us that
\begin{equation}
\label{D}
\sqrt{2q}P_q + 6HqG_q = D\mathrm{e}^{3N} \ ,
\end{equation}
where $D$ is an integration constant which has mass dimension 2.
If $q$ has a constant value as an initial condition, the Lagrangian (\ref{L2}) is also constant
and the universe evolves as a de Sitter spacetime, $H=const$. In such situation, the constant $D$
must be 0 because the derivative term in the field equation (\ref{f3}) vanishes. In what follows we will discuss reconstruction for the case when $q$ does {\it not} have a constant value and $D \neq 0$. For a formulation of the case $D=0$, see the Appendix.
%

Using (\ref{D}) and the background equations, we can write all quantities  $q(N)$, $H(N)$, and their derivatives with respect to $N$ as
\begin{align}
\label{P2}
   &P = -3\Pl^2H^2+D\sqrt{2q}\mathrm{e}^{3N} \ , \\
   \label{G2}
   &q'G_q= G'= \frac{2\Pl^2H'}{\sqrt{2q(N)}} - \frac{D\mathrm{e}^{3N}}{H} \ , 
\end{align}
and the quantities ${\cal F}_S$ and ${\cal G}_S$ are now
\begin{equation}
\label{ }
{\cal F}_S= {\cal F}_S(q,q',q'';N),\quad {\cal G}_S= {\cal G}_S(q,q';N)\ .
\end{equation}
If we provide explicit functions for $H(N)$ and $c_s=\sqrt{{\cal F}_S/{\cal G}_S}$, as in the case of $P(X)-V(\phi)$, we have here to solve a nonlinear second order differential equation on $q(N)$, i.e.
$c_s^2(N) = {\cal F}_S(q,q',q'';N)/{\cal G}_S(q,q';N)$.
Then, after obtaining two independent solutions of the differential equation, say $q_1(N)$ and $q_2(N)$, we invert them as $N_1(q)$ and $N_2(q)$. And substituting either of them, $N_1(q)$ or $N_2(q)$, into
(\ref{P2}) and (\ref{G2}), after integration we will get the explicit functions
 $P(q)$ and $G(q)$. Inserting now $q$ into $X$, we will finally obtain the so-called reconstructed Lagrangian.

However, two difficulties can be envisaged in this procedure, in spite of the fact that one can perform it step by step, in principle. First, it may generally be difficult to solve the nonlinear second order differential equation analytically to find the two solutions explicitly, and hence we may not be able to obtain analytic functions for the Lagrangian to be reconstructed. Second, in the case when we do have a explicit function for $c_s(N)$, we can ensure the subluminality of the propagation speed of the scalar field, but it is generically very unclear whether the stability condition (\ref{SC}) is satisfied or not. To confirm stability, we need to obtain rather complex forms for ${\cal F}_S$ and ${\cal G}_S$ using the solution
$q_1(N)$ or $q_2(N)$ after solving the differential equation on $q(N)$. Therefore, this is usually not a good way at all to provide a explicit function for $c_s(N)$ directly.

To confirm stability, it is more reasonable to tune ${\cal F}_S$ and ${\cal G}_S$ as they satisfy the stability condition (\ref{SC}), instead of giving a explicit function for $c_s(N)$ directly. In order to impose the stability condition (\ref{SC}) on models in a more convenient way, we first derive a relation between $H'$ and a product made of the undetermined function $G(q)$, instead of $c_s(N)$, as
\begin{equation}
\label{IN}
\Pl^2H' = -\sqrt{2q^3}{G_q} \ .
\end{equation}
%
In this case, the expression (\ref{G2}) turns into a {\it first} order differential equation on $q(N)$, namely
\begin{equation}
\label{q}
D\sqrt{2q}\e^{3N} = \Pl^2HH'\left( 2+ \frac{q'}{q}\right) \ .
\end{equation}
If we now provide a explicit function for $H(N)$, we have more chances to solve this differential equation without problems. This equation is still a nonlinear equation, but now it is first order, after setting the relation (\ref{IN}), and thus much easier to solve, in general, than the second order one. Furthermore,
the quantities ${\cal F}_S$ and ${\cal G}_S$ become
\begin{align}
\label{}
    &{\cal F}_S = \frac{\Pl^2\epsilon'}{(1+\epsilon)^2}, \quad
    {\cal G}_S = \frac{3\Pl^2\epsilon}{1+\epsilon}\left( 2\frac{q}{q'}+1\right) \ ,
\end{align}
where we should use the solution of (\ref{q}) for $q/q'$ in ${\cal G}_S$.
These expressions for ${\cal F}_S$ and ${\cal G}_S$ are relatively concise and thus we can constrain models to satisfy the stability condition quite easily. The stability condition (\ref{SC}) yields in our case
\begin{equation}
\label{SC2}
\epsilon'>0,\quad \epsilon(1+\epsilon) \left( 2\frac{q}{q'}+1\right)-\epsilon' \geq 0 \ .
\end{equation}
After we find a proper function for $H(N)$ which satisfies (\ref{SC2}) and the observational constraints (\ref{obs}), we invert $q(N)$ as $N(q)$, substitute it into (\ref{P2}) and (\ref{G2}) after
integration on $N$, and insert back $q$ into $X$, to finally obtain the reconstructed Lagrangian:
\begin{align}
\label{P2}
{\cal L} &= P(X) - G(X)g^{\mu\nu}\nabla_\mu \nabla_\nu \phi \ , \nonumber \\
    P(X)&= -3\Pl^2H^2(N(X)) + D\sqrt{2X}\e^{3N(X)} \ ,  \\
    \label{G2}
    G(X)&= \left. \int  \partial N\left( \frac{2\Pl^2H'}{\sqrt{2q(N)}} - \frac{D\e^{3N}}{H}\right)
                      \right|_{N=N(X)} \ .
\end{align}
It is guaranteed that from the Lagrangian created in this way, the function $H(N)$ we had originally provided in (\ref{q}) will be derived, and thus the resulting system will be stable. We note that the formalism we have discussed here is just one way to construct a stable system. We could assign a different relation related $c_s$, instead of (\ref{IN}), to determine the unknown functions $P(X)$ and $G(X)$. The reason we use the relation (\ref{IN}) is because it results in the tractable expressions
(\ref{SC2}) for the stability condition. Below, we will construct a specific model of G-inflation using the above formalism.

\subsubsection{Example 2}
As a concrete model of reconstruction of G-inflation, we consider the following one
\begin{equation}
\label{H2}
H^2(N) = H_0^2\mathrm{e}^{2\gamma N} + \Lambda^2 \ ,
\end{equation}
where $H_0$ and $\Lambda$ are constant with dimensions of mass, while $\gamma$ is a dimensionless constant.
We can solve the equation (\ref{q}) for this model and the solution is
\begin{equation}
\label{q2}
q(N) = \hat{D}^2\mathrm{e}^{(4\gamma-6)N} \,,\quad \hat{D}^2 \equiv \frac{8M_\mathrm{Pl}^4H_0^4}{D^2}\gamma^2(\gamma-1)^2 \ .
\end{equation}
Here,
\begin{align}
\label{ }
\epsilon &= \frac{\gamma}{1+x}, \quad \epsilon' = \frac{2\gamma^2 x}{(1+x)^2}, \nonumber \\
x&\equiv \frac{M^2}{H_0^2\mathrm{e}^{2\gamma N}}>0\ ,
\end{align}
and the stability condition reads
\begin{equation}
\label{SC3}
\frac{2\gamma^2 x}{(1+\gamma +x)^2} >0, \quad
\frac{(\gamma^2 +\gamma x + \gamma)\left(\frac{\gamma-1}{\gamma-\frac{3}{2}}\right)
- 2\gamma^2x}{(1+x+\gamma)^2} \geq 0 \ .
\end{equation}
If
\begin{equation}
\label{SC4}
0< \gamma \leq \frac{2-\sqrt{2}}{2}\quad  \text{or} \quad \frac{3}{2}\leq \gamma \leq \frac{2+\sqrt{2}}{2} \ ,
\end{equation}
the conditions above are always satisfied and the model is stable.
On the other hand, if we restrict to values of the period $x \ll 1$, the parameter region $\gamma \lesssim -1$ is also stable, although in this case the universe is in the phantom phase, where the null energy condition (NEC) is
violated: $H'<0\ (\dot H >0)$. However, $x$ should grow exponentially when $N$ increases  provided $\gamma$ is negative, and it would become much larger than 1 at very early times. Therefore, we have to set the initial condition such that $x \ll 1$ in order to obtain a stable NEC violating universe.

The values of ($f_s,\ g_s$) for this model are
\begin{equation}
\label{ }
f_s = \frac{2\gamma^2+2\gamma-2\gamma x}{1+\gamma+x}, \quad g_s = -\frac{2\gamma x}{1+\gamma+x}\ ,
\end{equation}
and the values of the observables (\ref{obs1})-(\ref{obs3}),
\begin{align}
\label{tilt2}
n_s-1 &= -\frac{2\gamma}{1+x} -\frac{3\gamma^2+3\gamma-\gamma x}{1+\gamma+x} \ , \\
r&= \frac{32\gamma^2x}{(1+\gamma+x)^{5/2}}\sqrt{\frac{(2\gamma^2-3\gamma)x}{3(\gamma-1)}} \ ,  \\
\alpha_s&= -\frac{2\gamma}{(1+x)^2}+\frac{12\gamma^2x}{(1+\gamma+x)^2} \ .
\end{align}
For simplicity, we shall consider  the cases $x\ll 1$ or $x\gg 1$ at $N=60$ only.
If we consider the case $x\gg 1$ at $N=60$ and discard all the other terms without $x$, the spectral index becomes
\begin{equation}
\label{ }
n_s -1 \sim \gamma \ . \nonumber
\end{equation}
From the observational data  (\ref{obs}) we have that $\gamma\sim-0.032$.
However, this value of $\gamma$ is excluded by the stability condition (\ref{SC4}), and thus we cannot obtain any stable inflationary model in the region $x\gg 1$ with the Hubble parameter (\ref{H2}) and the relation (\ref{IN}).
On the other hand, if $x\ll 1$ at $N=60$ and we ignore the terms proportional to $x$, the spectral index (\ref{tilt2}) becomes
\begin{equation}
\label{ }
n_s-1 \sim -5\gamma \ , \nonumber
\end{equation}
and the observational value for spectral index in (\ref{obs}) constrains $\gamma$ to be
\begin{equation}
\label{gamma}
0.00494 \lesssim \gamma \lesssim 0.00746 \ .
\end{equation}
This parameter region of $\gamma$ can indeed satisfy the stability condition (\ref{SC4}) and also other observational constraints in (\ref{obs}), as
\begin{align}
\label{}
    r&\sim 32\gamma^{5/2}x^{3/2} \ll 0.0002  \ , \nonumber \\
    \alpha_s &\sim -2\gamma,\quad -0.01492\lesssim \alpha_s \lesssim -0.00988 \ , \nonumber
\end{align}
Thus, we find values for the parameter $\gamma$ consistent with the observations and ensuring the stability of the model when $x\ll 1$. The tensor to scalar ratio is very much suppressed and this model has little amount of primordial gravitational waves in the consistent region of $\gamma$, (\ref{gamma}).

Hereafter we will consider  the case $x\ll 1$ and $\gamma= 0.006$ only. It is important to note that $\gamma$ is determined to be positive and, hence, the NEC violating
universe (in which $\gamma$ has a negative value) is excluded by the Planck observations.

In the case $x\ll 1$, the Hubble parameter is dominated by the first term on the right hand side of (\ref{H2}) and that term is responsible for
the energy of inflation. Thus, we can evaluate the values of $H_0$ and $\Lambda$ using the observational value for the scalar spectrum from Planck, as
\begin{align}
\label{H3}
&{\cal P}_\zeta \sim \left. \frac{H_0^2\e^{2\gamma N}}{8\pi^2c_s{\cal F}_S}\right|_{N=60} \ , \nonumber \\
&H_0^2\left ( \frac{H_0}{\Lambda}\right)^3 \sim 2.1\times10^{25}\text{GeV}^2  \gg H_0^2\ ,
\end{align}
where we have used the approximations ${\cal F}_S\sim 2\Pl^2\gamma^2 x$ and ${\cal G}_S\sim 2\Pl^2\gamma$.
(\ref{H3}) yields  $H_0 \ll 10^{13}\text{GeV}$, and the energy scale of inflation is relatively smaller than the typical energy scale for the potential driven inflation, as in Example 1. If we set $H_0= 10^{10}\text{GeV}$,
we can determine $\Lambda$ as $\Lambda=1.7\times 10^8 \text{GeV}$ from (\ref{H3}).

Now, we reconstruct the Lagrangian by substituting (\ref{H2}) and (\ref{q2}) and its inversion $N=N(q)$ into (\ref{P2}) and (\ref{G2}), with the result
\begin{align}
\label{}
{\cal L} &= P(X) - G(X)g^{\mu\nu}\nabla_\mu \nabla_\nu \phi \ , \nonumber \\
    P(X)&= \Pl^2H_0^2(4\gamma^2-4\gamma -3)( X/\hat{D}^2)^{\frac{1}{2\gamma-3}} -3\Pl^2\Lambda^2  \ ,\\
    G(X)&= \frac{\sqrt{2}(3-2\gamma)\Pl^2H_0}{\hat{D}}\sqrt{(X/\hat{D}^2)^{\frac{1}{2\gamma-3}}+\Lambda^2/H_0^2} \nonumber \\
    &\quad\cdot \sum_{n=0}^{249}\left(-\frac{\Lambda^2}{H_0^2}\right)^n \binom{249}{n} \frac{\left( (X/\hat {D}^2)^{\frac{1}{2\gamma-3}}+\Lambda^2/H_0^2\right)^{249-n}}{499-2n} \ ,
\end{align}
for $\gamma = 0.006$. Note that the quantity $\hat{D}$ does never affect the evolution of the system since  we can remove it from the Lagrangian by rescaling the scalar field as $\hat{\phi} =\phi/\hat{D}$, similarly as in Example 1. The Lagrangian obtained here does not include a canonical kinetic term and has a very complicated form. In general, the reconstructed Lagrangian does not always show the familiar form which includes a canonical kinetic term plus irrelevant terms. However, this Lagrangian will certainly produce observables which are fully consistent with the observational values and, moreover, the cosmological model will be stable.

We should stress again that ensuring the consistency with observations and an easy determination of the stability of the model are two of the main advantages in favor of using the reconstruction formalism over other methods. In exchange for these consistency and stability, one has to be ready to make sometimes sacrifices in terms of the simplicity of the Lagrangian. However, taking the lesson from effective field theories, the complexity of our Lagrangian can be explained in terms of the existence of an unknown (more fundamental) high energy theory which becomes valid at an energy scale much higher than the one corresponding to our `effective field theory', and we do not have to worry so much about the complexity of our Lagrangian, necessarily.
Again, the model we have shown here is just an example, and we are in principle able to reconstruct a Lagrangian which has a canonical kinetic term by using a different $H(N)$. The important thing to be noted is that we have presented in detail a full-fledged  formalism of reconstruction which can be applied to any function of $H(N)$.
\section{Conclusions}
\label{Conclusions}
We have developed in this paper a reconstruction formalism for two different types of Lagrangian, namely $P(X)-V(\phi)$ and
 G-inflation, with which we can make the models consistent with the observational data from the Planck mission and evaluate the stability of the models easily. We have reconstructed two specific examples for each type of Lagrangian which are safe from any instability and satisfy the observational constraints for the spectral index, the tensor to scalar ratio, and for the running of the spectral index.
For the case of $P(X)-V(\phi)$, although it exhibits a quadratic potential for $\phi$, we were able to obtain a model consistent with the observations by adjusting the sound speed $c_s$ to be smaller than $1$, namely $c_s=\sqrt{1/3}$. On the other hand, for the case of the G-inflation, we have been able to construct a stable model in a way which is  much easier than with other methods, all thanks to the reconstruction formalism we have developed in the paper.

We insist that the reconstruction formalism we have put up here is a very general procedure which can be applied, in principle, to any model if both $H(N)$ and a quantity related to $c_s(N)$ are provided. It is well known by now that using this and other existing formalisms, one always ends up with a Lagrangian which is consistent with the astronomical observations when one inputs the explicit functions of $H(N)$ and of the quantity related to $c_s(N)$ from the observational values of the power spectrum, (\ref{PS}) and (\ref{SS}), and so on. However, after applying these methods the Lagrangian one obtains does not necessarily satisfy the stability condition, a condition that is generically quite difficult to impose. And if the Lagrangian does not satisfy the stability condition, this means that the theory cannot actually interpret the observations, and we can exclude it. In our case we impose this condition at an earlier stage of the procedure, so that it is guaranteed in the end results and, hence, we can judge easily whether the resulting theory is appropriate to describe inflation or not. Further, our formalism does not make use of the slow roll approximation, and for this reason it can also be employed with all families of models in which the scalar field does not evolve with slow rolling.

Finally, in this paper we have not considered the reheating process and we have also not included ordinary matter in the action. To include matter is quite important, not only for dealing with the reheating process but also in order to be able to construct dark energy models using this reconstruction formalism. We leave these issues for a future work.

\vspace{5mm}
\noindent {\bf Acknowledgments.}
The investigation of EE has been supported in part by MINECO (Spain), projects FIS2010-15640 and FIS2013-44881, and by the CPAN Consolider Ingenio Project.

\appendix
\section{Reconstruction of the exact de Sitter solution in G-inflation}

In the case of exact de Sitter background and constant velocity of the scalar field, i.e. $H=const.,\ q=const.$,
the parameter $D$ must be zero for consistent evolution, and all background equations become
algebraic equations of $q$. At this time, the background equations (\ref{bg1}) become
\begin{align}
\label{K3}
    &3\Pl^2H^2 = - K   \ ,\\
    \label{f4}
    &\sqrt{2q}K_q + 6H\sqrt{2q^3}G_q= 0\ .
\end{align}
We obtain a variation of $H(q)$ by $q$ from (\ref{f4}) as
\begin{equation}
\label{ }
\frac{\delta H}{\delta q} = -\left[ \frac{K_{qq}}{3\sqrt{2q}G_q} -\frac{K_q}{6\sqrt{2q^3}G_q}
-\frac{K_qG_{qq}}{3\sqrt{2q}G_q^2} \right]\ .
\end{equation}
Using this variation, the coefficients of the linear scalar perturbation become
\begin{align}
\label{}
    {\cal F}_S&= \frac{\Pl^4q}{\Theta^2}\left(-\frac{K_q}{3} + \frac{qK_q^2}{3K}\right) \ , \nonumber  \\
    {\cal G}_S&= \frac{\Pl^4q}{\Theta^2}\left(-K_q +2qK_{qq} - \frac{qK_q^2}{K} -2q\frac{K_q}{G_q}G_{qq}
                        \right)  \ .
\end{align}
This expression is the generalization of the corresponding expression in \cite{Kobayashi:2010cm}.
In order to obtain a reconstructed Lagrangian, we have to provide two inputs, as the functions of $q$, $H(q)$ and $c_s^2(q)$, instead of $H(N)$, and a quantity related with $c_s(N)$, as the stability condition (\ref{SC}) is satisfied. For example, if we provide
\begin{align}
\label{}
    &H(q) = \sqrt{\frac{1}{3\Pl^2}\left( q- \frac{q^2}{2M_{dS}^3\mu}\right)}  \ , \\
    \label{cs3}
    &c_s^2 = \frac{y(1-y)}{6(1-y/2)\{1-y+1/(1-y/2)\}} \ , \\
    &y\equiv \frac{q}{M_{dS}^3\mu}, \quad 0< y < 1\ , \nonumber
\end{align}
where $M_{dS}$ and $\mu$ are constants which have a mass dimension.
The function $K(X)$ is trivially reconstructed from (\ref{K3}), as
\begin{equation}
\label{ }
K(X) = -X + \frac{X^2}{2M_{dS}^3\mu} \ ,
\end{equation}
and we obtain the quantities ${\cal F}_S$ and ${\cal G}_S$ using $K(X=q)$, as
\begin{align}
\label{F3}
    &{\cal F}_S = \frac{\Pl^4q}{\Theta^2}\frac{y(1-y)}{6(1-y/2)}  \ , \\
    \label{G3}
    &{\cal G}_S = \frac{\Pl^4q}{\Theta^2}\left( 1-y + 1/(1-y/2)-2q\frac{G_{qq}}{G_q}\right) \ .
\end{align}
Comparing (\ref{F3}) and (\ref{G3}) with (\ref{cs3}), we find $G_{qq}=0$, and hence the reconstructed function of $G$ is
\begin{equation}
G(X) = \tilde{D}X \ ,
\end{equation}
 where $\tilde{D}$
is an integration constant. We ignore the second integration constant since eventually it becomes a total derivative term in the action. If we set $\tilde D = 1/M_{dS}^3$, the background equation (\ref{f4}) determines the value of $\mu$,
\begin{equation}
\label{ }
\frac{6\mu^2}{\Pl^2} = \frac{(1-y_c)^2}{y_c^2(1-y_c/2)},
\end{equation}
where $y_c$ is just a number which has the same value as $y$, but its variation with respect to $q$ is zero, $\delta y_c = 0$, while $\delta y = \delta q/M_{dS}^3\mu$. This result is completely consistent with \cite{Kobayashi:2010cm}.


\end{document}